\documentclass[fleqn,usenatbib,useAMS]{mnras}

\usepackage{graphicx}	
\usepackage{amsmath}	
\usepackage{amssymb}	
\usepackage{multicol}        
\usepackage{bm}		
\usepackage{pdflscape}	
\usepackage{adjustbox}
\usepackage{hyperref}
\usepackage{arydshln}
\usepackage{dblfloatfix} 
\usepackage{tablefootnote}
\usepackage{xcolor}
\usepackage[normalem]{ulem}

\usepackage[T1]{fontenc}
\usepackage{ae,aecompl}
\usepackage{etoolbox}
\usepackage{newtxtext,newtxmath}

\title[Unsupervised methods for transient discovery]{Unsupervised machine learning for transient discovery in Deeper, Wider, Faster light curves}

\author[S. Webb]{\hyperlink{https://orcid.org/0000-0003-2601-1472}{Sara Webb}$^{1,2}$\thanks{Contact e-mail: \href{mailto:webb.sara.a@gmail.com}{webb.sara.a@gmail.com}}, Michelle Lochner$^{3,4,5}$, \hyperlink{https://orcid.org/0000-0002-5788-9280}{Daniel Muthukrishna}$^{6}$, Jeff Cooke$^{1,2}$, 
\newauthor   Chris Flynn$^{1,2}$, Ashish Mahabal$^{7}$, \hyperlink{https://orcid.org/0000-0002-9575-5152}{Simon Goode}$^{1,2}$, \hyperlink{https://orcid.org/0000-0002-8977-1498}{Igor Andreoni}$^{7}$, \hyperlink{https://orcid.org/0000-0001-9227-8349}{Tyler Pritchard}$^{8}$ 
\newauthor and \hyperlink{https://orcid.org/0000-0003-1587-3931}{Timothy M. C. Abbott}$^{9}$.
\\
$^{1}$Centre for Astrophysics and Supercomputing, Swinburne University of Technology, Mail Number H29, PO Box 218, 31122, Hawthorn, VIC, Australia \\
$^{2}$ ARC Centre of Excellence for Gravitational Wave Discovery (OzGrav), Australia \\
$^{3}$ Department of Physics and Astronomy, University of the Western Cape, Bellville, Cape Town, 7535, South Africa \\
$^{4}$ African Institute of Mathematical Sciences, Muizenburg, Cape Town, 7950 \\
$^{5}$ South African Radio Astronomical Observatory, Observatory, Cape Town, 7295 \\
$^{6}$ Institute of Astronomy, University of Cambridge, Madingley Road, Cambridge CB3 0HA, UK \\
$^{7}$ Division of Physics, Mathematics and Astronomy, California Institute of Technology, Pasadena, CA 91125, USA\\
$^{8}$ Center for Cosmology and Particle Physics, New York University, New York, NY 10003 \\
$^{9}$ NOIRLab, Mid-Scale Observatories/Cerro Tololo Inter-American Observatory, Casilla 603, La Serena, Chile
}

\date{Last updated 2019 September 22}

\pubyear{2019}
\begin{document}
\label{firstpage}
\pagerange{\pageref{firstpage}--\pageref{lastpage}}
\maketitle

\begin{abstract}
Identification of anomalous light curves within time-domain surveys is often challenging. In addition, with the growing number of wide-field surveys and the volume of data produced  exceeding astronomers' ability for manual evaluation, outlier and anomaly detection is becoming vital for transient science. We present an unsupervised method for transient discovery using a clustering technique and the \textit{Astronomaly} package. As proof of concept, we evaluate 85 553 minute-cadenced light curves collected over two $\sim$1.5 hour periods as part of the \textit{Deeper, Wider, Faster program}, using two different telescope dithering strategies. By combining the clustering technique HDBSCAN with the isolation forest anomaly detection algorithm via the visual interface of \textit{Astronomaly}, we are able to rapidly isolate anomalous sources for further analysis. We successfully recover the known variable sources, across a range of catalogues from within the fields, and find a further 7 uncatalogued variables and two stellar flare events, including a rarely observed ultra fast flare ($\sim$5 minute) from a likely M-dwarf. 

\end{abstract}





\newpage

\section{Introduction}
\vspace{1cm}
In the era of large time-domain surveys, classification and discovery of transient sources is becoming reliant on machine classification to handle the associated large amounts of data. Current ground based surveys such as the Zwicky Transient Facility \citep[ZTF, ][]{ Bellm2019PASP, Graham2019}, Dark Energy Survey  \citep{DES2016} and the All Sky Automated Survey for Supernovae \citep{ Shappee2014} are able to scan thousands of square degrees continuously, which amounts to petabytes of data annually, and recently the Panoramic Survey Telescope and Rapid Response System Survey \citep{Stubbs_2010, Chambers2016} delivered the first petabyte scale optical data release.  Space-based time-domain missions have provided unprecedented volumes of photometry, light curves, and proper motions for Galactic sources, with \textit{Kepler} \citep{Borucki2010} $\&$ \textit{K2} \citep{Howell2014} targeting $\sim$400,000+ individual stars, \textit{TESS} \citep{Stassun2018} is expected to target at least 200,000 sources producing light curves for each source, and Gaia has already released almost 2 billion sources.  Overcoming the mining challenges of these increasing amounts of data to not only identify and catalogue the multitude of known transient types but to make discoveries of new or anomalous sources is paramount to the success of future large transient surveys and time-domain science.
\subsection{Supervised Learning}
\indent Supervised machine learning has already been utilised extensively by several surveys and teams in astronomy for identification of variable stars and quasi-stellar objects from light curves via multivariate Gaussian mixture models, random forest classifiers, support vector machines, or Bayesian neural networks \citep[][]{Debosscher2007, richards2011,  Kim_2011, Pichara2012, Bloom_2012,Pichara2013, Kim2016, Mackenzie2016}. The literature aforementioned successfully shows the robustness of source classification while using the combination of supervised algorithms trained on extracted features. Features represent a set of measurable properties/characteristics of the light curves being studied (discussed in further detail in \ref{subsec:Features}). The most common features used in earlier works are available within the python package `\textit{FATS}' by \cite{Nun2015}.

Classification of non-folded light curves of extragalactic transient sources has also been explored, moving away from selecting the class of the object by fitting analytical templates built from a set of known sources \citep[][]{richards2011, Karpenka2012, Lochner2016, Narayan_2018, moller2016}.While these techniques work well for catalogues of light curves, they cannot easily be applied to real-time data.

Real-time classification of supernovae by \cite{Muthukrishna2019} and \cite{moller2019} has shown the effectiveness of deep recurrent neural networks, without the need to rely on extracting computationally expensive features of the input data. 

\subsection{Unsupervised Learning}
Even with machine learning advances in astronomy, mining data for unknown or anomalous events is relatively unexplored, as the majority of current algorithms require training data sets of known events.  \cite{Mackenzie2016} developed an unsupervised feature learning algorithm that takes subsections of variable star light curves to cluster and use as features to train a linear support vector machine.  This work eliminates the need for traditional feature extraction, limiting the computing time and biases associated with feature selection. Only limited work into actual transient classification or anomaly detection via unsupervised means has been performed within time domain astronomy.

\cite{Valenzuela2018} performed unsupervised clustering of variable star light curves by creating variability trees using the $k$-medoids clustering algorithm of fragmented light curves. This method offers a novel and computationally fast approach to data exploration but is again limited by the need for known light curve examples for similarity searches.  To identify \textit{Kepler} data outliers for visual inspection, \cite{Giles2019} performed light curve clustering using Density-Based Spatial Clustering of Applications with Noise (DBSCAN).  They report the successful extraction of the known anomalous Boyajian's star via their method; however they identified that the DBSCAN assumption of constant density clusters is a limitation.
It should be noted that the overwhelming majority of work performed to date on light curve classification by machine learning has used 30 minute to several day cadence, including folded light curves.

\cite{2017arXiv170906257M} presents another  approach to light curve classification, by reducing the time series data to two-dimensional representation in order to classify them using deep learning techniques. This approach maps the change and magnitude over time to create a visual representation of the light curve as an image to be used in the deep learning process. This method presents an alternate approach of unsupervised learning for time-series classification without the need for feature extraction.

\subsection{Anomaly detection in fast cadenced surveys}
Currently, the majority of wide field optical surveys explore a limited region of the luminosity-timescale phase space, with an average cadence of hours to days between visits to fields, with only a few programs exploring the phase space shorter than 1-hour cadence \citep[see,][]{Becker2004, Rykoff2005,  Lipunov2007, Rau2009, Berger2013, Burdge2019,Richmond2020PASJ}.  What is largely unexplored by these surveys is the phase space of transient events occurring on seconds-to-minutes time scales. There are several events expected to occur on these timescales, and understanding both the events and the general nature of the fastest transients in the Universe is crucial for understanding the transient Universe as a whole. For example, the upcoming Rubin Observatory Legacy Survey of Space and Time (LSST) is predicted to generate nearly 10 million transient alerts each night.  As such, it will be invaluable to quickly and meaningfully quantify the expected large volume of short timescale events to help assist in follow-up priority assignment \citep{LSST2009}. To do so, the astronomical community will rely heavily on the use of brokers and their integrated algorithms serving alert streams. Current brokers, which include ALeRCE\footnote{https://github.com/alercebroker}, ANTARES\footnote{https://antares.noao.edu}, Lasair\footnote{https://lasair.roe.ac.uk}, and MARS\footnote{https://mars.lco.global} are already in use on the nightly ZTF stream, successfully identifying known extragalactic and galactic transient and variable events. However identifying anomalous events can prove challenging with pre-trained algorithms, especially within the rarely explored fast timescales (seconds-to-minutes). 

The multi-wavelength Deeper, Wider, Faster (DWF) program offers the ability to explore optical transient events with the depth and cadence required to enable the quantification and characterisation of Galactic and extragalactic variable and fast transient rates for current and upcoming large-area searches and surveys and to similar depths as 4m - 8m class telescopes. Such as gravitational wave counterpart searches, the Rubin Observatory LSST survey, and others. This work presents our effort to explore the DWF optical data for anomalous light curves without the restrictions of prior assumptions or expectations. 

As our literature review highlights, the vast majority of work to date on machine learning for transient classification and identification has relied on pre-existing understanding of longer duration variable and transient time-series behaviour. In this work, we demonstrate an unsupervised method to aid in the discovery of both known and poorly understood transients on the timescales of seconds-to-minutes. 

The paper is organised as follows: A brief introduction to the DWF program is presented in Section \ref{subsec:DWF}, two DWF data gathering strategies and the data in Section \ref{subsec:data}.  We present our multifaceted anomaly detection approach in Section \ref{subsec:Methodology} and our proof of concept results in Section \ref{subsec:Results}.  We conclude by presenting our overall outcomes in Section \ref{subsec:Conclusion}.

\section{The Deeper, Wider, Faster Program}
\label{subsec:DWF}

Several new and exciting astronomical fast transient events have been discovered in recent decades and the progenitors and physical mechanisms behind many of them are still poorly known (e.g., Fast Radio Bursts (FRBs)), supernova shock breakouts, Fast-Evolving Luminous Transients (FELTs) and other rapidly evolving extragalactic events \citep[for example:][]{Lorimer2007, Garnavich2016, Prentice2018,Perley2018, Rest2018NatAs}.  What has limited our ability to detect and understand these events is the capability to gather data in short, regular time intervals before, during and after the events; as well as over a range of wavelengths.  The DWF program \citep[][]{Meade2017, vohl2017, andreoni, andreoni2, andreoniandcooke} has been designed with these challenges specifically in mind, constructing an all wavelength and simultaneous observational program of over 70 facilities to date.  DWF takes a `proactive' approach to transient astronomy, with coordinated simultaneous wide-field fast-cadenced multi-wavelength observations of target fields taken continuously over 1--3 hour periods, capturing data  before, during and after the transient events. The optical data collected during the simultaneous observations is processed in near real-time to quickly identify candidates requiring the use of rapid Target of Opportunity (ToO) observations. 

DWF unites the worlds most sensitive facilities with large fields of view in the optical --- the Dark Energy Camera  \citep[DECam,][]{Flaugher2015AJ} on the Cerro Tololo Inter-American Observatory (CTIO) Blanco 4-m telescope in Chile and Hyper-SurprimeCam  \citep[HSC,][]{Miyazaki2017} on the Subaru 8-m telescope in Hawaii --- taking continuous 20-30 second exposures. Using this strategy, DWF is able to explore a region of luminosity phase space rarely explored by many traditional surveys \citep[see][]{Andreoni2019A}. From the real-time data processing, DWF can quickly identify candidates and coordinate rapid-response and long-term follow-up observations of transient candidates. DWF began in 2014 and since its inception has had two pilot runs and seven operational runs \citep[see,][Cooke et al., in prep]{andreoniandcooke}. 

The unique design of DWF allows exploration of transients on the seconds-to-hours timescales, providing further understanding into known classes of fast transients, events theorised to occur on these timescales, and very early detections of slower-evolving events (see Section \ref{subsec:data} for observation specifics).  Using either DECam or HSC, the deep optical component of DWF can explore a region of parameter space not yet reached by previous transient surveys.  Note that, although DWF collects simultaneous fast-cadenced data across all wavelengths, radio through gamma-ray, from multiple facilities, we will only focus on DECam optical data here. 
Work by \cite{Andreoni2019A} utilised the unique DWF data and \textit{`Mary'}, our transient difference image discovery pipeline, to detect both galactic and extragalactic transients on the minute timescales. In this paper, we examine light curves generated purely from science images (i.e., without image subtraction) for all sources in our chosen fields, and explore the ability to identify known and unknown transient and variable sources through the use of unsupervised machine learning.  By examining every source light curve through an unsupervised algorithm, we aim to not only distinguish clear source separations in feature space, but identify and classify unknown and outlying sources to comprehensively explore fast transient events and source variability on the seconds-to-hours timescales. 

\section{Data}
\label{subsec:data}
We use fast cadenced data collected during DWF runs using DECam.   We collect 20-second, continuous imaging of targeted fields, acquired in a single band, the \textit{`g'} filter.  We choose the continuous use of the \textit{`g'} filter to maximize depth with DECam, reaching $\sim$0.5 magnitudes deeper in comparison to the other filters in dark time. The expected limiting magnitude in `g' band is m(AB) $\sim$23, for an average seeing of 1.0 arcseconds and airmass of 1.5 (relatively high airmass due to the field constraints of observing simultaneously with multiple facilities). For this work, the DECam images are post-processed through the NOAO High-Performance Pipeline System \citep{Valdes2007, Swaters2007, Scott2007} and then transferred to the OzSTAR supercomputer at Swinburne University of Technology for our data analysis.  The DECam 62 CCD mosaic is separated into individual fits files for each extension.  Each CCD is processed separately for source extraction using SExtractor \citep{Bertin1996} and all source magnitudes are corrected for exposure time and magnitude offsets against the SkyMapper Data Release 2 catalogue \citep{Bertin2010, Onken2019}.  A master list is compiled by cross-matching all extracted sources from each CCD, over all exposures within 0.5 arcsecond radius between source centroids into one catalogue of source positions.  This master catalogue is used to create light curves for each source, replacing any non-detections per single exposure with the CCD exposure detection upper limit represented in the light curve.  

To date, DWF has targeted 20 separate fields, each observed typically for 6 consecutive nights, and has accumulated over 1 million source detections. In this work, we analyse light curves from two separate fields for only one night each, observed using two different observing strategies.  In Section \ref{sec:J04-55}, we analyse data collected from the DWF `J04-55 field' on 18 December 2015, using a field centre of RA:04:10:00.0 $\&$ DEC: $-$55:00:00.0.  The continuous 20 second exposures were collected over a 90 minute period, using a \textit{stare}' observational
strategy (i.e., pointing at the same coordinates with no small field dithering between exposures).  In Section \ref{sec:antlia}, we analyse data gathered over an 80 minute period of continuous 20 second exposures centred on the `\textit{Antlia field}' RA: 10:30:00.0 $\&$ DEC: $-$35:20:00.0 on the Antlia cluster of galaxies.  These data were collected on 06 February 2017 and utilised a five point dithering strategy at the beginning middle and end of the observation, while staring in between. In these data, we explore the contribution  of telescope dithering to the false positive rate of anomaly detection in Section \ref{sec:antlia}.

\section{Methodology}
\label{subsec:Methodology}
We use the following methodology: (1) feature extraction, (2) clustering, (3) t-SNE visual representation, (4) anomaly ranking and visualisation with \textit{Astronomaly}. We use feature extraction to find a low dimensional representation of the data, clustering to eliminate large clusters of ordinary objects and instrumental effects and isolate possible interesting transients, anomaly detection to rank these remaining objects by "abnormality" and finally \textit{Astronomaly} to visually explore the detected anomalies. Note that all stages are performed on nightly light curves with an average cadence of $\sim$60-68 seconds between light curve points, accounting for both the 20\,s exposure and 40\,s CCD readout time, CCD clear and rest. We utilize python for all stages, using the following packages \textit{scikit-learn, hdbscan, FATS, astropy, numpy, pandas} and \textit{matplotlib}  \citep{scikit-learn, McInnes2017, Nun2015, astropy:2018, Oliphant2015, mckinney2010data, Hunter:2007}.  

\subsection{Features}
\label{subsec:Features}
As the number of data points differ for different light curves, we extract a uniform set of features to (a) reduce the dimensionality, and (b) allow for direct comparison between light curves that may be on different time scales with different sampling properties. To represent our unique fast-cadenced data, we use a mixture of normalised features developed and used primarily for the identification of variable stars and quasi-stellar objects. We performed principle component analysis on the features and selected those that corresponded to large eigenvalues. The majority of our features are taken from work by \cite{richards2011}, which were used to classify variable stars from sparse and noisy time-series data. We use only the features not restricted explicitly to folded light curves or periodic sources. Some examples of the features used are amplitudes, standard deviation, linear trend, maximum slope, etc. In addition to these, we used the stellar variability detection features, H$_{1}$(amplitudes), R$_{21}$ (the 2nd to 1st amplitude ratio), and R$_{31}$ (the 3rd to 1st amplitude ratio) which are focused around Fourier decomposition. The remaining features were taken from work in quasi-stellar object selection, these being auto-correlation length, consecutive points, variability index and Stetson KAC as used by \cite{Kim_2011} and mean, $\sigma$ and $\tau$ taken from a continuous autoregressive model fitted to our data from \cite{Pichara2012}.  We extract 25 unique features from each light curve using mostly using \textit{FATS} and some in-house routines. Full details and sources for the features used are shown in Appendix \ref{tab:featuresappendix1}.
 
In this work we run feature extraction in parallel on the OzSTAR supercomputer at Swinburne University. We utilize the Intel Gold 6140 18-core processors on OzSTAR, achieving a feature extraction speed of  $\sim$ 110 seconds per 1000 light curves when processed serially. 

\subsection{HDBSCAN}
The focus of this paper is to use machine learning to analyse and cluster our light curves. We choose to use Hierarchical Density-Based Spatial Clustering of Applications with Noise \citep[HDBSCAN\footnote{https://hdbscan.readthedocs.io/en/latest}, ][]{McInnes2017}. The theoretical method behind this algorithm was first proposed by \cite{Campello2013}.  HDBSCAN takes the approach of Density-Based Spatial Clustering of Applications with Noise (DBSCAN) and converts it into a hierarchical clustering algorithm by varying the value of epsilon ($\epsilon$) to identify clusters of varying densities  \citep[for further details see][]{McInnes2017}. 

To better understand how HDBSCAN works, we first outline the original DBSCAN algorithm by \cite{Ester1996}.  DBSCAN performs nearest neighbour searches in a given feature space to determine clusters of over-densities, points closely related in distance, and identify outlier points that exist in low density regions as noise.  DBSCAN requires two parameters,  $\epsilon$, which represents the radius of the neighbourhood search and a minimum number of points ($minPts$), which must exist in a neighbourhood to constitute a dense region.  What has limited the use of DBSCAN in the past is the inability to vary $\epsilon$ in a given data set, requiring clusters to have similar densities.  However, HDBSCAN can take in a minimum cluster size parameter which eliminates the need for a single value of $\epsilon$ when determining clusters from a dendrogram, adjusting of $\epsilon$ as it explores clusters of varying densities.  

After several preliminary tests combining the different distance metrics and varying minimum cluster sizes to evaluate cluster purity and uniformity, we opted to require a minimum cluster size of 5 and to use a Euclidean distance metric for its intrinsic ability to calculate the shortest distance between points.  
We aim to create as many distinct clusters in our feature space as the algorithm will allow to limit the outliers to very low density regions. 

\subsection{t-SNE}
\label{subsec:t-SNE}
To help visualise the clustering of objects in our high dimensional feature space, we use the t-distributed Stochastic Neighbor Embedding (t-SNE) algorithm developed by \cite{vanDerMaaten2008}.  The t-SNE algorithm uses the same Euclidean distance metric to measure the proximity of all features in higher-dimensional space. It converts these distances to probabilities using a Gaussian distribution.  A similarity matrix of the probabilities is stored for the higher-dimensional space, and the feature space is then collapsed down to 2 or 3 dimensions, depending on the user's choice, where the Euclidean distance is calculated once again using a t-distribution to assign probabilities and saved as a second similarity matrix.  The two distributions are then minimized using the sum of Kullback-Leibler divergence of all data points using a gradient descent method to return a 2 dimensional representation of the distance of data in our feature space. It is important to note that due to the stochastic nature of t-SNE, it is used here only for visualisation and not cluster identification. We note here that t-SNE was performed for the entirety of our data sets, using the OzStar\footnote{https://supercomputing.swin.edu.au/ozstar/} computing nodes as well as on a personal machine with 8 GB ram and a 4.0 GHz quad-core Intel Core i7.
We acknowledge that for future work the use of Uniform Manifold Approximation and Projection for Dimension Reduction  \citep[UMAP,][]{2018arXiv180203426M} is a promising method for dimensionality  reduction, however in this work we were unable to use UMAP due to computational issues and we deemed t-SNE to be sufficient.

\subsection{Astronomaly}
To find the most anomalous light curves, in each cluster, we use the python package \textit{Astronomaly}\footnote{https://github.com/MichelleLochner/astronomaly} (Lochner $\&$ Bassett in prep) which is comprised of a python back end and JavaScript front end to easily explore the data via a locally hosted web interface (for further details see Appendix \ref{fig:Astronom}). \textit{Astronomaly} is a flexible framework, designed to detect anomalies within astronomical images or light curves using any of a variety of anomaly detection algorithms. Here we use the scikit-learn implementation of isolation forest \citep{isolationf} available in \textit{Astronomaly}. Each cluster of light curves identified by HDBSCAN was saved in individual data frames containing each light curve's features.  

Using \textit{Astronomaly}, each cluster's light curve's where evaluated independently, feeding both their features and original light curve file into the back end of the package.

The isolation forest then works to isolate each light curve by recursively generating partitions, creating a tree structure ultimately segregating each light curve point into nodes. Each node either contains one individual data point, or several data points all with the same feature value. 

The web interface GUI allows the user to visually inspect the highest ranking anomalous light curves (as measured by the isolation forest algorithm), as well as explore the interactive t-SNE plot to probe the lower dimensional cluster space. To enable more rapid visualisation, for this work we limit \textit{Astronomaly} to present only the 2000 most anomalously ranked light curves in the GUI interface.

\textit{Astronomaly} serves two purposes in this work. The first is easy visualisation of the data in the clusters. Each cluster is analysed individually and the interactive t-SNE plot allows the user to quickly determine if the objects in the cluster do indeed look similar. The data can then be further vetted using the ranked anomaly system. The most anomalous objects within the cluster will appear first and hence should be the objects that are least likely to actually belong to that cluster. Thus the effectiveness of the clustering can be quickly evaluated without the need for exhaustive study of every single light curve in the cluster.

The second reason we use \textit{Astronomaly} is to identify anomalous sources in the ``unclustered'' group. With the same ranking system, the most interesting sources (and also instrumental effects) should appear early in the list allowing quick identification. It is critical to note that while this dataset is still small enough to manually investigate every object (especially with Astronomaly's visual interface), for datasets consisting of millions of light curves this would simply not be possible and the automated ranking becomes much more important to allow rapid discovery of anomalous sources.

\section{Results}
\label{subsec:Results}
\subsection{DWF J04-55 field - \textit{No dithering} observational strategy}
\label{sec:J04-55}
We present the results of our unsupervised method applied to light curves over a 90-minute observation of the DWF `J04-55 field' using DECam in stare mode (the telescope tracked the same field centre coordinates for the duration of the observations). It is important to acknowledge that small movements of the telescope may still be present due to telescope guiding, shutter movements and small pointing shifts.  A total of 89 images were acquired, with 23 199 sources, as identified in the J04-55 field from the 5-night master source list, as having greater then 3 detections ($N_{det} > 3$) for feature extraction.

 \begin{table}
    \centering
    \begin{tabular}{lccr}
    \hline 
     \textbf{Description of} & \textbf{Cluster} &  \textbf{$\#$ of} & \textbf{$\%$ of}   \\
     \textbf{light curves} &  \textbf{ID} & \textbf{Light Curves} & \textbf{Sources} \\
    \hline 
    \hline 
    Faint sources &   \\
     at detection  & Cluster 0  &   8   & 0.03$\%$    \\
     threshold \\\hdashline[1pt/1pt]
    Sources near   \\
    CCD edge &  Cluster 1  &   144   & 0.62$\%$    \\\hdashline[1pt/1pt]
    Steady light curves & Cluster 2& 22909 & $>$98.7$\%$ \\ \hdashline[1pt/1pt]
    Real and &  \\
    photometrically  & Unclustered   &   138   & $<$ 0.59$\%$       \\
    affected light curves \\
    \hline
    \end{tabular}
    \caption{The details of each of the three clusters identified by the HDBSCAN algorithm.  The description of the light curves refers to both the light curve and information gathered from individual cutouts of the detection images.  unclustered represents light curves unable to be identified to a cluster. }
    \label{tab:J04-55cluster}
\end{table}

\begin{figure*}
    \centering
    \includegraphics[height=12cm]{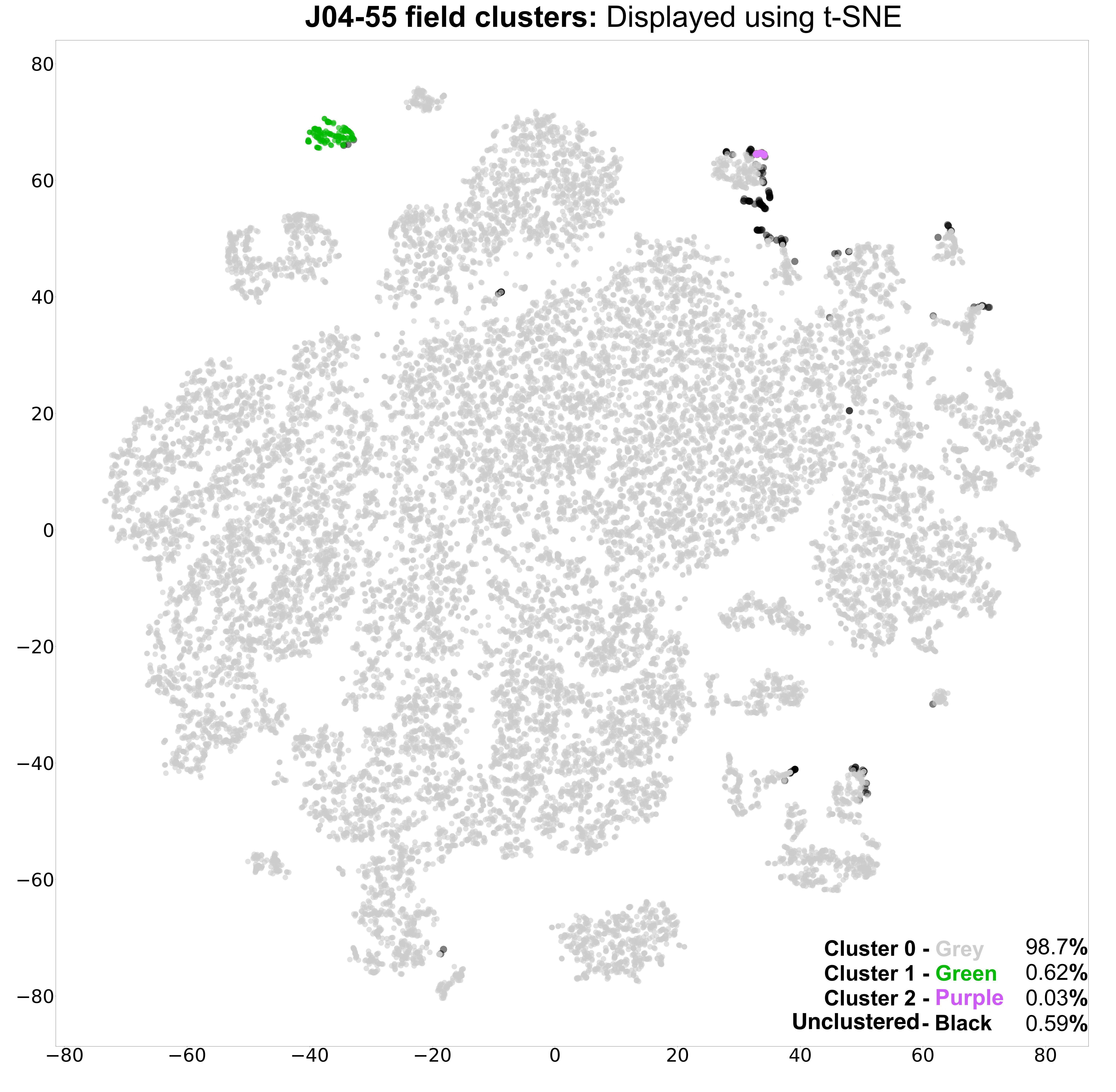}
    \caption{Feature space of the 25 features of the 23,199 light curves of the `J04-55 field' collapsed down to 2 dimensions using t-SNE with the clusters labelled in Table \ref{tab:J04-55cluster} and coloured accordingly. It is important to note 1) that the axis values within a t-SNE are not physically meaningful and hence not labeled, and 2) that the t-SNE algorithm works by adapting its own notion of distance to regional density variations in the higher dimensional data.  As a result, t-SNE naturally expands dense clusters and contracts sparse ones when collapsed as shown, and this can make some structure within the t-SNE plot appear more significant than it is.
    }
    \label{fig:TSNEJ04-55}
\end{figure*}

\subsubsection{Clusters}
A total of three clusters were identified using HDBSCAN, as shown in Table \ref{tab:J04-55cluster}. Cluster 2 dominates, containing $98.7\%$ of light curves in the field. Inspection showed that this cluster overwhelmingly contained sources which were unchanging in magnitude, consisting of both stars and galaxies. In such a short time-scale observation, we expect that the majority of sources will be assigned to a single cluster in this manner. The two remaining clusters identify faint sources only breaching the detection threshold a few times during the 90 minutes, and sources near, or on, the edges of CCDs which have caused unusual/anomalous light curves. A visual representation of the clusters in feature space can be seen in Figure \ref{fig:TSNEJ04-55}.

\begin{figure*}
    \centering
    \includegraphics[width=16.5cm]{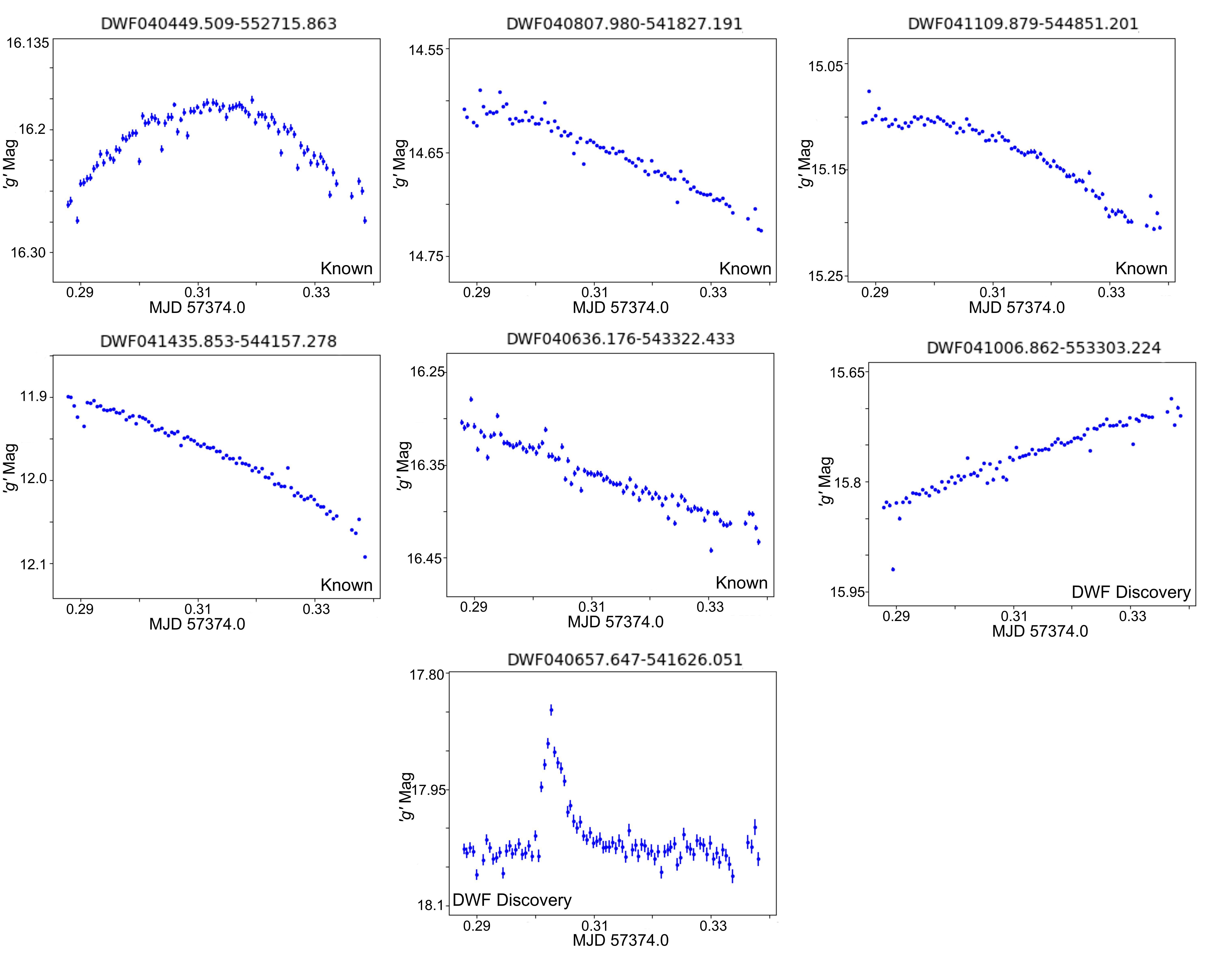}
    \caption{Four previously known and two newly discovered variable/transient sources present in the unclustered noise within the J04-55 field analysis.
    }
    \label{fig:subJ04-55noiseplots}
\end{figure*}

\begin{table*}
    \centering
    \begin{tabular}{ccccc}
    \hline
    \textbf{Field}  &\textbf{DWF ID}   & \textbf{Catalogued ID} & \textbf{Type$^{a}$} & \textbf{Period (Days)$^{b}$} \\ 
    \hline
    \hline
    J04-55 & DWF040449.509-552715.863 & ASASSN-V J040449.48-552715.9 & W Ursae Majoris & 0.27 \\
    J04-55&DWF040807.980-541827.191    & ASASSN-V J040807.97-541827.2  & W Ursae Majoris & 0.35  \\
    J04-55&DWF041109.879-544851.201    & SSS J041109.9-544851        & W Ursae Majoris & 0.32  \\
    J04-55&DWF041435.853-544157.278    & ASAS J041436-5441.9           & Contact Binary                            & 0.45 \\
    J04-55&DWF040636.176-543322.433    & DES 11110400160736          & RR Lyrae                                  & 0.59   \\
    \textbf{J04-55} & \textbf{DWF041006.862-553303.224}    & \textbf{Discovered in this work}       & \textbf{Slow pulsating B. }                            &  \textbf{-}   \\
    \textbf{J04-55} & \textbf{DWF040657.647-541626.051}    & \textbf{Discovered in this work} & \textbf{Flare event on RR Lyrae}  &    \textbf{0.86}   \\
    \hline
    \end{tabular}
    
    \raggedright $^{a}$ For previously catalogued sources, type is identified by catalogue, if newly discovered  source, type approximated from CMD position (see Appendix \ref{fig:CMD}).  \\
    \raggedright $^{b}$ For previously catalogued sources the period is taken from the discovery survey, if newly discovered source period is not known.  \\
    \raggedright $^{c}$ Absolute G-Band magnitude as calculated using \textit{GAIA} parallax information.
    \caption{Sources identified showing variability in J04-55 and Antlia fields. Note: lines in bold indicate discoveries in this work. }
    \label{tab:j04-55sourcetable}
\end{table*}

\subsubsection{Variable/Transient Sources}
\label{sec:J04-55unclusered}
A total of 138 light curves remained unclustered (referred to as \textit{noise} by HDBSCAN, shown in black on Figure \ref{fig:TSNEJ04-55}). The unclustered light curves represent those which have a significant distance from identified clusters and represent the outliers in the data. It is these outliers which are variable and transient sources in the field. The light curve of each was visually inspected (in order of anomaly score) using the Astronomaly package and variable sources were cross-matched to existing catalogs to check for known variability (mainly the International Variable Star Index (VSX) catalogue \citep{Watson2006}, identified RR lyrae stars from the Dark Energy Survey (DES) \cite{Stringer2019}, and the Catalina Surveys Southern Periodic Variable Star Catalogue \citep{Drake2017}). For newly discovered sources showing variability, locations on a Colour-Magnitude Diagram (CMD) were calculated using GAIA data release 2 parallax and photometric information \citep{gaiadr2photom2018, gaiadr2para2018}. The CMD positions were then overlaid on the variability CMDs presented in work by \cite{2019A&A...623A.110G} and shown in Appendix \ref{fig:CMD} as green triangles. After evaluation with Astronomaly, it was determined that the majority of the light curves were indeed anomalous in structure, however caused by instrumental and observational effects. The false positives represented sources on the edges of CCDs or those teetering on the detection threshold. However we did identify 6 sources of continuous variability, 5 of which have been previously catalogued, with the remaining variable source discovered by this work. In addition to the variable stars, a  stochastic classical flare event was also identified. Source IDs, name, coordinates, known catalog ID (if available) and period are shown in Table \ref{tab:j04-55sourcetable}, and the light curves are shown in Figure \ref{fig:subJ04-55noiseplots}.

\subsubsection{Validating the completeness for J04-55 field}
\label{sec:4hrval}
To confirm the effectiveness of our unsupervised clustering we used several methods to verify that all variable sources in the field were identified. First we retrieved all known variable sources from the VSX catalog. We found 13 catalogued variable sources within DECam's CCD footprint. Five of the known variable sources were recovered as anomalies in this work (see Table \ref{tab:j04-55sourcetable}), and three were below our detection threshold for the vast majority of exposures. The remaining five did not show significant variability over the $\sim$90 minute period and were subsequently clustered in the grouping of steady light curves. \ These four sources have catalogued periodicities much longer than 90 minutes (See Appendix \ref{tab:known} for their details.)
Secondly, Astronomaly was used to display the 2000 light curves ranked most anomalous via the isolation forest algorithm over the identified clusters. After visual inspection, no additional variable light curves were found. Through these evaluations we confirm that our methods successfully retrieve most, if not all, varying or transient sources present in the field during our observations.

\subsection{DWF J10-35 (or Antlia) field - \textit{Dithering} observational strategy}
\label{sec:antlia}
\begin{table*}
    \centering
    \begin{tabular}{cllcrl}
    \hline 
    \textbf{Sub}  & \textbf{Description of light curves} & \textbf{Cluster IDs} &  \textbf{$\#$ of} & \textbf{$\%$ of}  & \textbf{colour in}  \\
    \textbf{Group} & & & \textbf{Light Curves} & \textbf{Sources} & \textbf{t-SNE } \\
    \hline 
    \hline 
    G1 & Steady light curves       &  36     &  58279     &  93.5 $\%$     & Grey       \\
    G2 & Variable sources        &   1    &  6  &   $<$ 0.01$\%$    &   Cyan   \\
    G3 & Faint sources at detection threshold           &   33, 34, 35           &   23    &   $<$ 0.01$\%$    & Red \\ 
    G4 & Only detected on five point dithers      &   0, 3, 4, 21, 22, 23, 27, 28, 32  &  111     &  $<$ 0.2$\%$      & Orange \\ 
    G5 & Photometric correction issues on first 5 dither points     &  5, 6, 7, 9, 10, 11, 12, 13, 18   &   266    &  $<$ 0.45$\%$     & Blue \\
    G6 & Sources near edge of CCD resulting in  dimming and brightening    &   2, 14, 17, 24,26, 29     &    1176   &  1.88$\%$      &   Purple   \\
    G7 & One or more detections affected by cosmic rays, pixel faults, etc      &  31     &  5     & $<$ 0.01$\%$        &  Green\\
    G8 & Other photometric correction issues eg. Blended sources.   &  8, 15, 16, 19,20, 25, 30     &   319    &  $<$ 0.6$\%$      &  Pink \\
    UC & Contains a mixture of real variables and light curves affected     &       &       &       & \\
    &  by many of the identified photometic concerns outlined above  & -1 / unclustered  &  2169 & 3.48$\%$ & Black  \\
    \hline
    \end{tabular}
    \caption{The nine sub groupings of light curve types as identified in the Antlia field. }
    \label{tab:antliasubgrouping}
\end{table*}

Through the uniqueness of the DWF program, novel and nontraditional observing strategies have been implemented dependent on the strategies of the facilities performing simultaneous observations and the overall goals of the observing program. Here we confirm that our unsupervised analysis is able to successfully identify and quantify both real astrophysical anomalies, and those caused due to an observing stragtegy with relatively large dithers ($\sim$60 arcsec) designed to move the telescope sufficiently to fill the DECam CCD gaps evenly with 5 dithers.  We chose a DWF field where observations were a mixture of five point dithers, and continuous stares over an $\sim$80 minute period.  Dithering within surveys is often crucial to fill CCD chip gaps and gather photometic information of all sources in the field. Dithering in this manner results in partial light curves for sources in the chip gaps that are missed during the stare mode observations. Here we evaluate the `J10-35' field, which we will refer to as the  Antlia field, as the 3 deg$^2$ field is centred on the Antlia galaxy cluster.  The observations contained three, five point dithers during the beginning, middle and end of the observations.  

Using observations taken on the 06 February 2017, a total of 70 348 sources were identified in the Antlia field from the 5-night master source list.  Of these, 62 354 light curves met our pipeline criterion of having $N_{det} > 3$. Over the $\sim$80 minute observation period.

The same 25 features chosen previously were extracted from each of the 62,354 light curves and a total of 37 clusters were identified through the HDBSCAN clustering algorithm, as well as a group of unclustered light curves that did not satisfy the distance requirements to join the identified clusters (see Appendix \ref{tab:antliaclustersexpanded} for individual cluster information).  It is immediately apparent that a significantly higher number of clusters were identified throughout these data in comparison to the previous J04-55 field results in Section \ref{sec:J04-55}, for which we only find four clusters.  The increase in clusters is due to characteristics introduced into the light curves from photometric issues caused mainly by the dithering strategy and the tip/tilt motion when using the hexapod\footnote{The hexapod mechanism is a set of six pneumatically driven pistons that actuate to precisely align the optical elements between exposures.} on DECam. Below, we outline the usefulness of these clusters in identifying and quantifying transient classifications.

\begin{figure*}
    \centering
    \includegraphics[width=16cm]{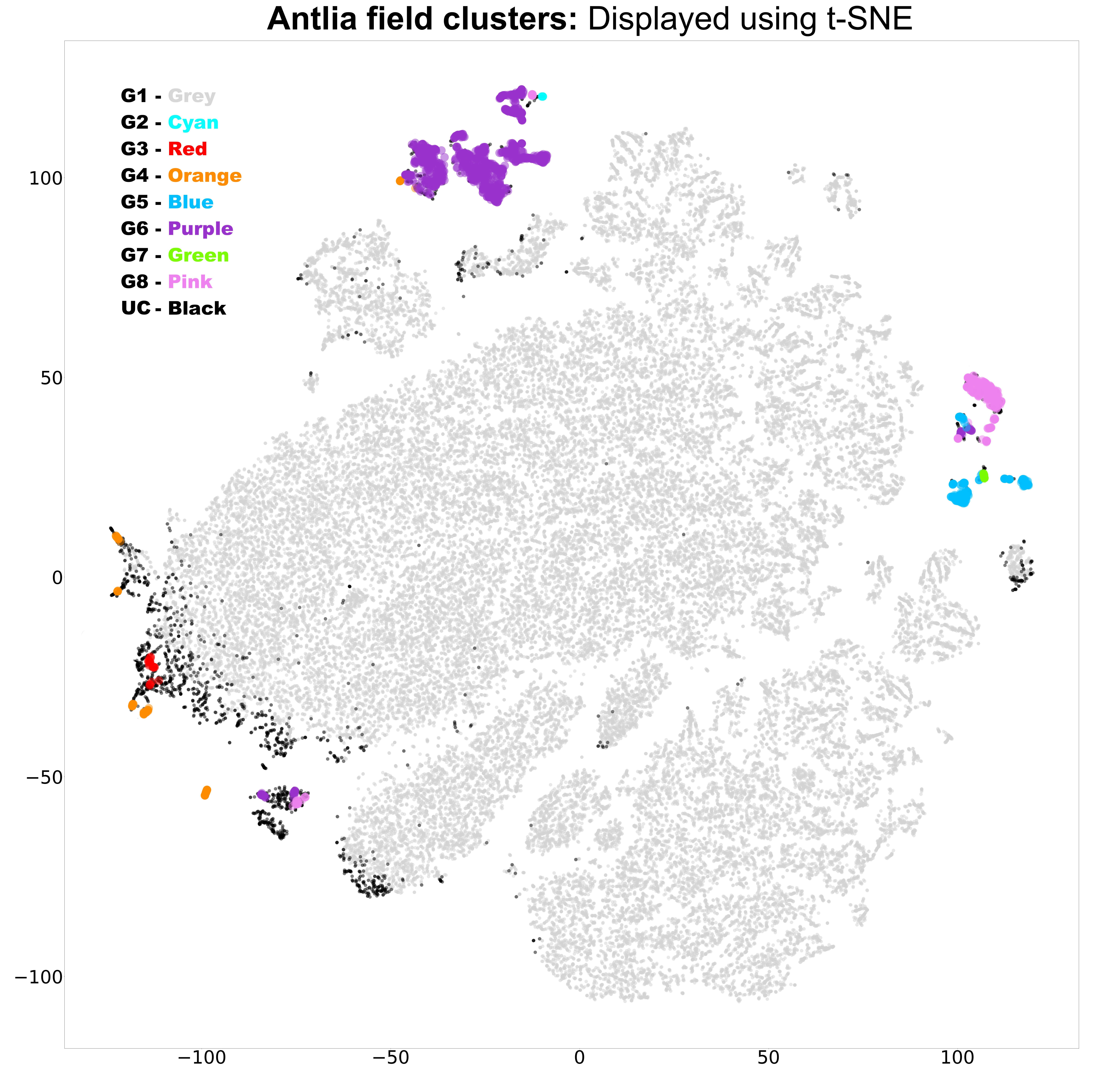}
    \caption{Feature space of the 25 features of the 62,354 light curves of the Antlia field collapsed down to 2 dimensions using t-SNE.  The sub groupings as outlined in Table \ref{tab:antliasubgrouping} are coloured accordingly.  It is important to note that t-SNE algorithm works by adapting its known notion of distance to regional density variations in the higher dimensional data, as a result t-SNE naturally expands dense clusters and contracts sparse ones when collapsed. 
    }
    \label{fig:TSNEantlia}
\end{figure*}
\begin{table*}
    \centering
    \begin{tabular}{ccccc}
    \hline
    \textbf{Field}  &\textbf{DWF ID}   & \textbf{Catalogued ID} & \textbf{Probable Type$^{a}$} & \textbf{Period (Days)$^{b}$}  \\ 
    \hline
    \hline
    Antlia & DWF102919.102-355133.303    & SSS J102919.0-355133 & Spotted Star & 0.34  \\
    Antlia & DWF102938.901-345415.969    & SSS J102938.8-345416 & W Ursae  Majoris & 0.27 \\
    Antlia & DWF103105.927-360744.003    & SSS J103105.8-360742 & W Ursae Majoris  & 0.44  \\
    \textbf{Antlia} & \textbf{DWF102552.421-354418.436}    & \textbf{Discovered in this work} & \textbf{$\delta$ Scuti or $\gamma$ Doradus} & \textbf{-}\\
    \textbf{Antlia} & \textbf{DWF103240.961-344522.875}    & \textbf{Discovered in this work} &  \textbf{-} & \textbf{-} \\
   \textbf{Antlia} & \textbf{DWF103147.030-354553.653}    & \textbf{Discovered in this work} &  \textbf{-} & \textbf{-} \\ 
    \hline
    \end{tabular}
    
    \raggedright $^{a}$ For previously catalogued sources, type is identified by catalogue, if newly discovered  source, type approximated from CMD position (see Appendix \ref{fig:CMD}).  \\
    \raggedright $^{b}$ For previously catalogued sources the period is taken from the discovery survey, if newly discovered source period is not known.  \\
    \raggedright $^{c}$ Absolute G-Band magnitude as calculated using \textit{GAIA} parallax information.
    \caption{Sources identified showing variability in J04-55 and Antlia fields. Note: lines in bold indicate discoveries in this work. }
    
    \label{tab:antlia1sourcetable}
\end{table*}

\subsubsection{Cluster Sub Groupings}
\label{sec:clustersub}
The 37 clusters can be broken down into eight sub groups of clusters, including the unclustered grouping, shown in Table \ref{tab:antliasubgrouping}.  Visual inspection of randomly selected, if not all for the smaller groupings, source fits images over time were used to determine the sub groupings. The majority of clusters fall into the subgroups of photometic anomalies caused by telescope dithering, photometric correction issues or, less frequently, by CCD artifacts/cosmic rays. However two sub groupings are of interest, variable sources (G2), and the light curves that were unable to be clustered with HDBSCAN (UC within Figure \ref{fig:TSNEantlia} and Table \ref{tab:antliasubgrouping}). The variable sources identified in G2 are discussed further in Section \ref{sec:antliagroupvar}. 

\begin{figure*}
    \centering
    \includegraphics[width=15.5cm]{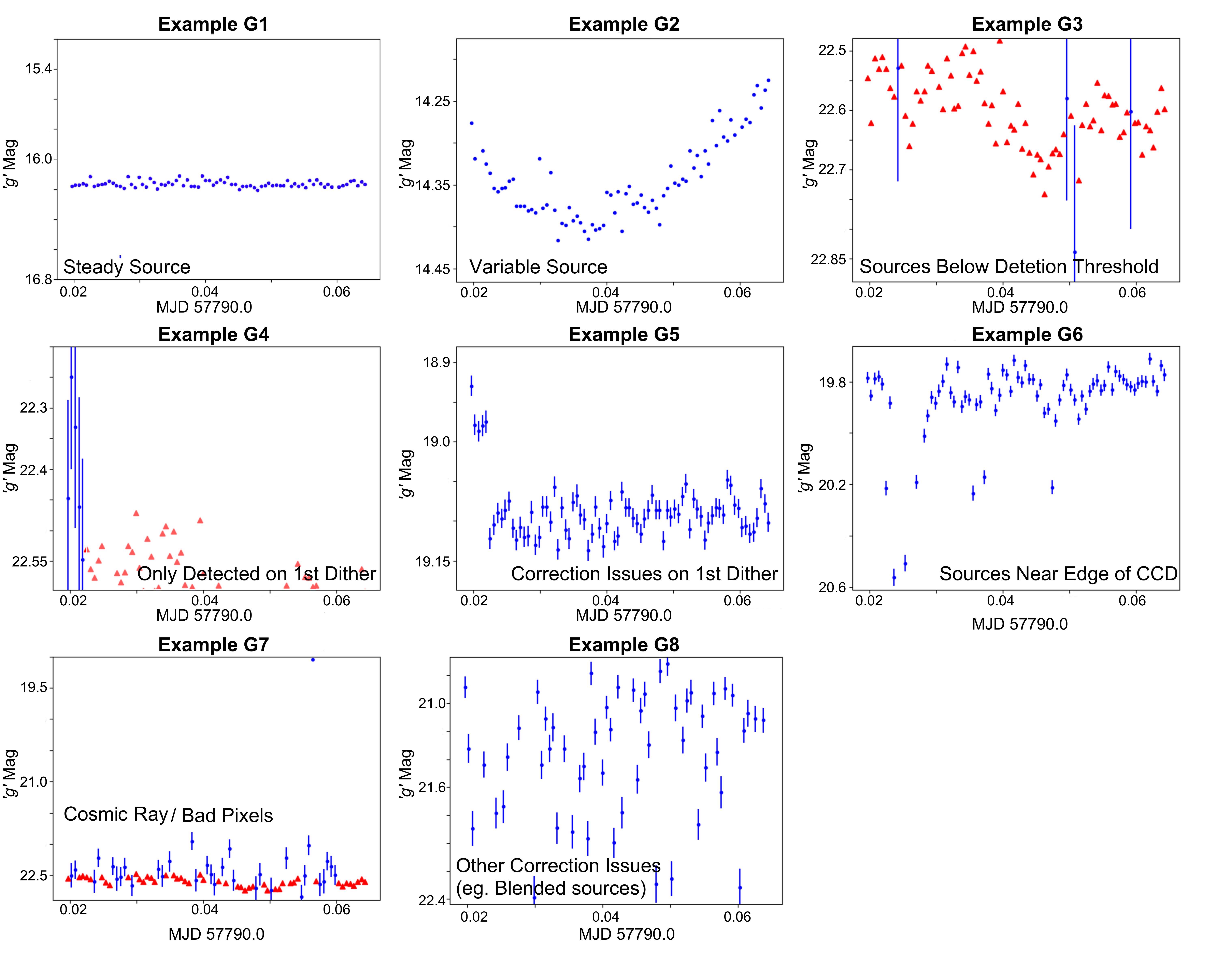}
    \caption{Antlia field examples of typical light curves present in each of the sub groupings. The blue points represent source detections the red triangles represent the limiting magnitudes of the exposures and are only present in the light curves when sources are not detected.
    }
    \label{fig:subgroups}
\end{figure*}

Representation of the clusters in feature space can been seen in Figure~\ref{fig:TSNEantlia} where the feature space has been reduced into 2 dimensions using t-SNE. The figure clearly shows the feature space dominated by one main cluster of non-varying light curves (number 36, sub group G1), which is unsurprising, as we expect the majority of sources in the field to be unchanging over the minutes-to-hours time scales. Figure~\ref{fig:TSNEantlia} further illustrates the grouping of clusters with related light curves by highlighting the sub groups of light curve properties and their causes as outlined in Table \ref{tab:antliasubgrouping}.  Example light curves of each of the sub groups are shown in Figure \ref{fig:subgroups}.

From the sub grouping of clusters, we are able to meaningfully quantify the light curves for this field: finding that 93.5$\%$ are grouped into one cluster, of steady light curves, while $\sim$ 2.0$\%$ of light curves were affected by telescope dithering and/or the use of the hexapod on the DECam instrument, and 0.39$\%$ of light curves had photometric correction issues over the first 5 exposures (of the 80) due to the initial five point dither pattern and change in standard stars used for correction on certain CCDs.

\subsubsection{Sub groups identifying Variable Sources}
\label{sec:antliagroupvar}
The algorithm identified one cluster containing sources of true astrophysical variability, described in sub grouping G2 in Table \ref{tab:antliasubgrouping}. These sources were cross-matched to several catalogs to check for known variability, as outlined in Section \ref{sec:J04-55unclusered}. In this group we identified 6 variable sources, 3 of which have been previously catalogued and 3 sources discovered by this work. Source IDs, name, coordinates, known catalog ID (if available) and period are shown in Table \ref{tab:antlia1sourcetable}.

\begin{figure*}
    \centering
    \includegraphics[width=16cm]{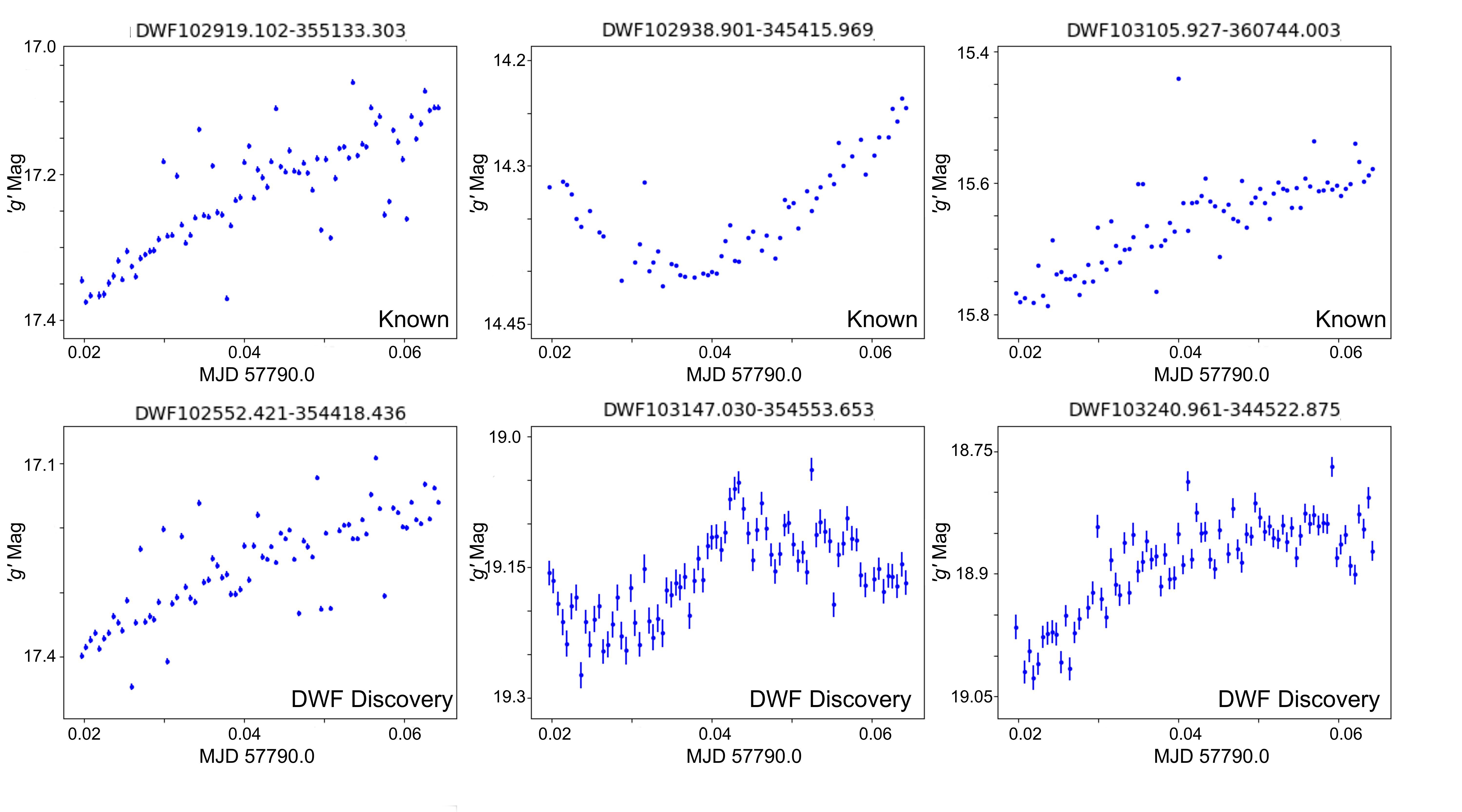}
    \caption{Three Previously known and three newly discovered variable sources as identified in sub group G2.}
    \label{fig:antlia_c1vars}
\end{figure*}

Of the 3 newly discovered sources in this sub grouping we are unable to unambiguously identify the variable types of two sources using the CMD in Appendix \ref{fig:CMD}. The CMD location of the reamining source was calculated using GAIA data release 2 parallax and photometric information \citep{gaiadr2photom2018, gaiadr2para2018}. The CMD position is overlaid on the variability CMDs presented in work by \cite{2019A&A...623A.110G} and subsequently used for likely type identification in \ref{tab:antliasubgrouping}. We are unable to confidently classify DWF103240.961-344522.875 in the CMD because of its large GAIA parallax uncertainty and, thus, absolute magnitude. On the other hand, DWF103147.030-354553.653 sits in an area where few pulsating objects are found, between main sequence stars and white dwarfs but where cataclysmic variables are common. A source in this region was shown by \cite{gaiadr2var2019} to be likely a cataclysmic variable (CV). The light curves for all 6 sources are presented in Figure \ref{fig:antlia_c1vars}.

\subsubsection{Variable/Transient Sources}
\label{sec:antliaunclustered}

A total of 2 169 light curves were unclustered by HDBSCAN and not assigned to a specific cluster in our analysis of the Antlia field. These light curves can be seen to sit along the outskirts of the main grouping of G1 in Figure~\ref{fig:TSNEantlia}, as well as occupying similar feature space to other identified clusters. It is these light curves which are of particular interest for rare transient and variable events, as we expect any unusual and unique light curves in comparison to the majority to be identified as noise via HDBSCAN. 

Two independent approaches were used to evaluate the unclustered light curves. The first was manual inspection of all 2 169 light curves and the second was anomaly detection and ranking using Astronomaly. This dual approach was taken to comparatively quantify the successful extraction of interesting anomalous light curves using \textit{Astronomaly's} inbuilt isolation forest anomaly ranking. Here, \textit{Astronomaly} was used to explore groupings of similar light curves through its inbuilt interactive t-SNE plot.

During our evaluation, sources within the unclustered grouping, were again cross-matched to VSX, DES and the Catalina Surveys Southern Periodic Variable Star Catalogue, to identify previous detections and classifications. The majority of the unclustered light curves were false positives caused by dithering affects on sources. However, amongst the false positives we identify 9 variable sources, 6 of which were previously catalogued by surveys, with the remaining 3 sources discovered in this work. We further discover an ultra fast flaring source, with positioning on the CMD suggesting the source is consistent with M dwarf flares. Optical flare events evolving on very short timescales (seconds-to-minutes) such as this have previously only been identified using 10 second cadence of NUV GALAX data by \cite{Brasseur2019ApJ}, uncovering a previously unexplored population of short duration of stellar flares. Source IDs, name, coordinates, known catalog ID (if available) and period are shown in Table \ref{tab:sourcetable}. The light curves for each of the sources are presented in Figure \ref{fig:antlia_noisevars}.  The newly discovered sources showing variability are overlaid on the CMD in Appendix \ref{fig:CMD} as purple triangles.

\begin{figure*}
    \centering
    \includegraphics[width=15.5cm]{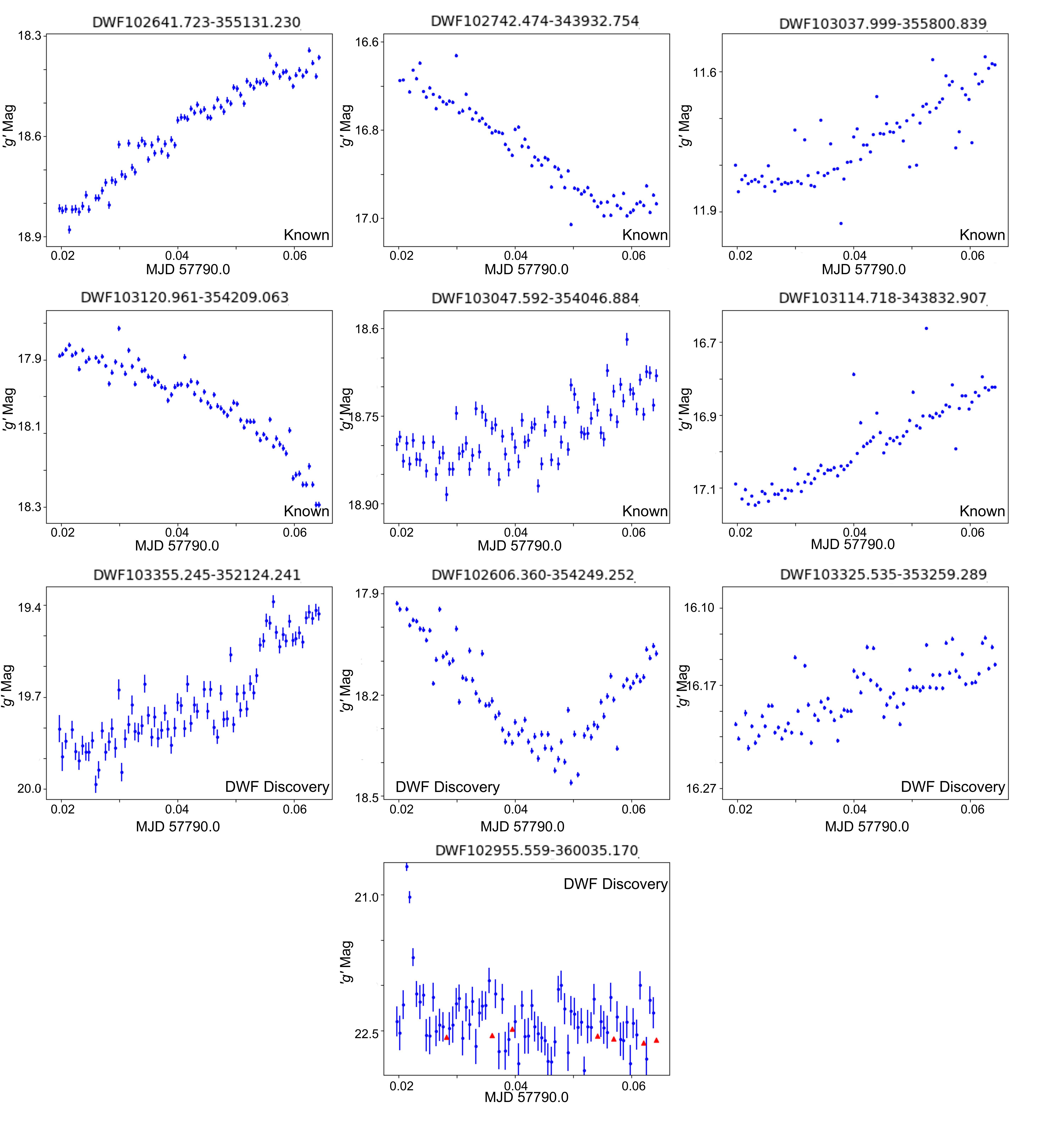}
    \caption{Size Previously known and four newly discovered variable sources as identified in the grouping of unclustered light curves. The blue points represent source detections, while the red triangles represent the limiting magnitudes of the exposures and are only present in the light curves when sources are not detected.}
    \label{fig:antlia_noisevars}
\end{figure*}

\begin{table*}
    \centering
    \begin{tabular}{cclcc}
    \hline
    \textbf{Field}  &\textbf{DWF ID}   & \textbf{Catalogued ID} & \textbf{Type$^{a}$} & \textbf{Period (Days)$^{b}$}  \\ 
    \hline
    \hline
    Antlia & DWF102641.723-355131.230    & SSS J102641.7-355130 & W Ursae Majoris & 0.29 \\
    Antlia & DWF102742.474-343932.754   & SSS J102742.4-343933 & W Ursae Majoris & 0.27  \\
    Antlia & DWF103120.961-354209.063   & SSS J103120.8-3542094 & W Ursae Majoris & 0.27 \\
    Antlia & DWF103037.999-355800.839   & ASAS J103038-3558.0   & $\beta$ Persei  & 0.72  \\
    Antlia & DWF103047.592-354046.884   & SSS J103047.5-354047  & RR Lyrae  &  0.31  \\
    Antlia & DWF103114.718-343832.907   & SSS J103114.5-343834. & RR Lyrae  & 0.33  \\
    \textbf{Antlia} & \textbf{DWF102606.360-354249.252}   & \textbf{Discovered in this work} & \textbf{UV Ceti or T Tauri} & \textbf{-}  \\ 
   \textbf{Antlia} & \textbf{DWF103355.245-352124.241}   & \textbf{Discovered in this work} & \textbf{T Tauri}     & \textbf{-}  \\
    \textbf{Antlia} & \textbf{DWF103325.535-353259.289}   & \textbf{Discovered in this work}  & \textbf{ $\gamma$ Doradus} & \textbf{-}  \\
    \textbf{Antlia} & \textbf{DWF102955.559-360035.170}   & \textbf{Discovered in this work} & \textbf{Ultra fast flare} & \textbf{-}  \\ 
    \hline
    \end{tabular}
    
    \raggedright $^{a}$ For previously catalogued sources, type is identified by catalogue, if newly discovered  source, type approximated from CMD position (see Appendix \ref{fig:CMD}).  \\
    \raggedright $^{b}$ For previously catalogued sources the period is taken from the discovery survey, if newly discovered source period is not known.  \\
    \raggedright $^{c}$ Absolute G-Band magnitude as calculated using \textit{GAIA} parallax information.
    \caption{Sources identified showing variability in J04-55 and Antlia fields. Note: lines in bold indicate discoveries in this work. }
    
    \label{tab:sourcetable}
\end{table*}

\subsubsection{Astronomaly Performance}

We utilised the large set of unclustered light curves identified in the Antlia field to test the abilities of \textit{Astronomaly} to present only the most astrophysically anomalous light curves to astronomers in a timely manner. \textit{Astronomaly} takes less than 2 minutes to process the features through the isolation forest algorithm and launch the interactive web GUI.

Using the \textit{Astronomaly} front end GUI to visually inspect each light curve in ranked order, we identified the nine variable sources within the  top 280 of 2000 highest ranking anomalous light curves taken from the grouping of unclustered sources and the ultra fast flare event was identified within the first 600. By using both clustering and \textit{Astronomaly} we were able to find all the anomalies in the first 0.9$\%$ of the over all Antlia data. This result highlights the possibility to significantly reduce the amount of time needed for light curve evaluation of anomalous events by astronomers, and will be continued to be utilised in the future analysis of DWF light curves.

A more recent version of \textit{Astronomaly} contains human-in-the-loop learning, designed specifically to deal with finding objects that are swamped by more anomalous points (according to the machine learning) but are actually more mundane objects.

\subsubsection{Validating the completeness for Antlia field}

Similar to Section \ref{sec:4hrval} we took several steps to verify all variable sources which were identified. Within a 1.5 degree radius of the field centre, 22 catalogued variable sources (with periods less than 1 day) existed in the VSX catalogue and within DECam's CCD footprint. Nine of the known variable sources were recovered as anomalies in this work, both being identified in the cluster of variables and within the unclustered grouping of most anomalous light curves, as explained in detail in Sections \ref{sec:clustersub} and \ref{sec:antliagroupvar}. Of the remaining sources, 6  did not show significant variability over the $\sim$80 minute period and were subsequently clustered in the grouping of steady light curves, consistent with their longer recorded periods (See Appendix \ref{tab:knownant} for full details). The remaining 7 were either below detection threshold, at saturation limits or photometrically affected by dithering and were clustered accordingly.  
\textit{Astronomaly} was used to display the top 2000 light curves (limited to 2000 light curves by \textit{Astronomaly} for the handling of the interactive t-SNE plot) ranked most anomalous via the isolated forest algorithm over the identified clusters. After visual inspection, no additional interesting light curves were found.  

\section{Conclusion}
\label{subsec:Conclusion}
Existing and future astronomical surveys are continuously pushing the bounds of the known transient universe, and the ability to efficiently probe a large number of light curves in a timely manner will become vital in the exploration of regions of previously known and unknown classes of events.  In this work, we have successfully shown the capability of unsupervised machine learning methods to rapidly and thoroughly explore fast cadenced data collected by transient surveys, using the DWF program as an example. By taking a two-step approach of both clustering and anomaly/outlier detection, we were able to identify 7 previously unidentified variable stars. We also identified two classes of stellar flares, one classical flare and one rapidly evolving flare, further demonstrating the effectiveness of our unsupervised methods and the unique capability of the DWF program. Notable is the speed of which this method can be performed.
Feature extraction takes $\sim$110 seconds per 1000 light curves and when run in parallel (on the OzSTAR supercomputer) can complete a set of 70,000 light curves in less then 15 minutes. The HDBSCAN clustering takes a further $\sim$2 minutes, and in total, a set of 70,000 light curves can be ready for human evaluation using \textit{Astronomaly} within 20 minutes. Both the speed and ease of use our method demonstrates the ability of unsupervised methods in meaningfully evaluating light curves to identify source variability. This method is well suited for the use on current and upcoming surveys for anomaly detection, for which hundreds of millions of light curves will inevitably be produced.
\\
\indent Finally, we stress that this work explores a small fraction of the full DWF data set, only 2 fields for 80-90 minutes each.  Future work will involve the evaluation of 250+ hours of data for 17 fields.  Moreover, as DWF runs typically occur over 6 consecutive nights, additional variable sources will be found over a range of phase durations when the data is analysed over the full run duration for the 2 fields explored here.  Furthermore, we plan to use this unsupervised method on light curves combined over multiple nights to search for long period variables, which would otherwise appear steady in single night light curves.



\section*{Acknowledgements}

We would  like  to  thank  the  organisers  of  the  2019  Kavli Summer Program in Astrophysics hosted at the University of California, Santa Cruz, without which this collaboration and work would not have been possible. The program was funded by the Kavli Foundation, The National Science Foundation,UC Santa Cruz, and the Simons Foundation. Part of this research was funded by the Australian Research Council Centre of Excellence for Gravitational Wave Discovery (OzGrav), CE170100004. We acknowledge the financial assistance of the National Research Foundation (NRF). Opinions expressed and conclusions arrived at, are those of the authors and are not necessarily to be attributed to the NRF.This work was partly supported by the GROWTH (Global Relay of Observatories Watching Transients Happen) project funded by the National Science Foundation under PIRE Grant No 1545949. This work has made use of data from the European Space Agency (ESA) mission Gaia (https://www.cosmos.esa. int/gaia), processed by the Gaia Data Processing and Analysis Consortium (DPAC, https://www.cosmos.esa. int/web/gaia/dpac/consortium). Funding for the DPAC has been provided by national institutions, in particular the institutions participating in the Gaia Multilateral Agreement. \\
\noindent
This project used data obtained with the Dark Energy Camera (DECam),
which was constructed by the Dark Energy Survey (DES) collaboration.
Funding for the DES Projects has been provided by 
the U.S. Department of Energy, 
the U.S. National Science Foundation, 
the Ministry of Science and Education of Spain, 
the Science and Technology Facilities Council of the United Kingdom, 
the Higher Education Funding Council for England, 
the National Center for Supercomputing Applications at the University of Illinois at Urbana-Champaign, 
the Kavli Institute of Cosmological Physics at the University of Chicago, 
the Center for Cosmology and Astro-Particle Physics at the Ohio State University, 
the Mitchell Institute for Fundamental Physics and Astronomy at Texas A\&M University, 
Financiadora de Estudos e Projetos, Funda{\c c}{\~a}o Carlos Chagas Filho de Amparo {\`a} Pesquisa do Estado do Rio de Janeiro, 
Conselho Nacional de Desenvolvimento Cient{\'i}fico e Tecnol{\'o}gico and the Minist{\'e}rio da Ci{\^e}ncia, Tecnologia e Inovac{\~a}o, 
the Deutsche Forschungsgemeinschaft, 
and the Collaborating Institutions in the Dark Energy Survey. 
The Collaborating Institutions are 
Argonne National Laboratory, 
the University of California at Santa Cruz, 
the University of Cambridge, 
Centro de Investigaciones En{\'e}rgeticas, Medioambientales y Tecnol{\'o}gicas-Madrid, 
the University of Chicago, 
University College London, 
the DES-Brazil Consortium, 
the University of Edinburgh, 
the Eidgen{\"o}ssische Technische Hoch\-schule (ETH) Z{\"u}rich, 
Fermi National Accelerator Laboratory, 
the University of Illinois at Urbana-Champaign, 
the Institut de Ci{\`e}ncies de l'Espai (IEEC/CSIC), 
the Institut de F{\'i}sica d'Altes Energies, 
Lawrence Berkeley National Laboratory, 
the Ludwig-Maximilians Universit{\"a}t M{\"u}nchen and the associated Excellence Cluster Universe, 
the University of Michigan, 
{the} National Optical Astronomy Observatory, 
the University of Nottingham, 
the Ohio State University, 
the OzDES Membership Consortium
the University of Pennsylvania, 
the University of Portsmouth, 
SLAC National Accelerator Laboratory, 
Stanford University, 
the University of Sussex, 
and Texas A\&M University.
\noindent
Based on observations at Cerro Tololo Inter-American Observatory, National Optical
Astronomy Observatory 
which is operated by the Association of
Universities for Research in Astronomy (AURA) under a cooperative agreement with the
National Science Foundation.

\section*{Data Availability}

The data underlying this article will be shared on reasonable request to the corresponding author.

\bibliographystyle{mnras}
\bibliography{bib.bib} 

\newpage
\appendix
\onecolumn
\section{Features}
\begin{table*}
    \centering
    \begin{adjustbox}{width=1\textwidth}
    \begin{tabular}{rlll}
    \hline
    \textbf{Feature} & \textbf{Description} & \textbf{Inputs} & \textbf{Refs}  \\
    \hline
    \hline
    Amplitudes & Half the difference between  & Magnitude & \cite{richards2011}\\
    & the median of the maximum 5$\%$ and the median \\
    & of the minimum 5$\%$ Magnitude.\\
     Auto correlation length & Length of linear dependence of a signal with &  Magnitude &  \cite{Kim_2011} \\ 
     & itself at two points in time \\ 
     Beyond1Std & Percentage of points beyond one &  Magnitude $\&$ Error &  \cite{richards2011}\\
     & standard deviation from the weighted mean \\
     CAR$_{mean}$ & The mean of a continuous time auto regressive & Magnitude, Time $\&$ Error & \cite{Pichara2012}\\
     & model using a stochastic differential equation \\
     CAR$_{\sigma}$ & The variability of the time series on &  Magnitude, Time $\&$ Error & \cite{Pichara2012} \\
     & time scales shorter than $\tau$ \\
     CAR$_{\tau}$& The variability amplitude of the & Magnitude,  Time $\&$ Error & \cite{Pichara2012}\\
     &  time series \\
      H$_1$ & Amplitude derived using the Fourier & Magnitude &   \cite{Kim2016} \\ 
      & decomposition \\
     Con & The number of three consecutive & Magnitude &   \cite{Kim_2011} \\
     & data points that are brighter or fainter then 2$\sigma$ \\
     & and normalized by N -2 \\
     Linear Trend & Slope of a linear fit to the light curve & Magnitude $\&$ Time &  \cite{richards2011}\\ \\
     MaxSlope & Maximum absolute magnitude slope between two & Magnitude $\&$ Time &  \cite{richards2011}\\
     & consecutive observations \\
     Mean & The mean magnitude & Magnitude &  \cite{Kim2014} \\ \\
     Mean Variance & the
ratio of the standard deviation &Magnitude &  \cite{Kim_2011}\\
& to the mean magnitude \\
     Median Absolute Deviation & The median discrepancy of the data & Magnitude &  \cite{richards2011}\\
     & from the median data \\
     Median Buffer Range Percentage & Fraction of photometric points &Magnitude &  \cite{richards2011}\\
     & with amplitude/10 of the median magnitude \\
     Pair Slope Trend & The fraction of increasing first differences &Magnitude & \cite{richards2011} \\
     &  minus the fraction of decreasing\\
     &  first differences \\
     Q31 & The difference between the 3rd & Magnitude &  \cite{Kim2014} \\
     & and 1st quarterlies \\
     R$_{21}$ & 2$^{nd}$ to 1$^{st}$ amplitude ratio derived & Magnitude &\cite{Kim2016}\\
     & using the Fourier decomposition \\
     R$_{31}$ & 3$^{rd}$ to 1$^{st}$ amplitude ratio derived & Magnitude& \cite{Kim2016}\\
     & using the Fourier decomposition \\
     Rcs & Range of cumulative sum &Magnitude & \cite{richards2011}\\
     \\
     Skew & The skewness of the sample & Magnitude & \cite{richards2011}\\ \\
     Slotted Auto Correlation  & Slotted auto correlation length & Magnitude $\&$ Time & \cite{Protopapas_2015}\\ 
     Function Length \\ \\
     Small Kurtosis & Small sample kurtosis of magnitudes &Magnitude & \cite{richards2011}\\ \\
     Standard Deviation & Standard deviation of the magnitudes &Magnitude & \cite{richards2011}\\ \\
     Stetson K$_{AC}$ & Stetson K applied to the slotted &Magnitude & \cite{Stetson1996, Kim_2011} \\ 
     & auto correlation function of the light curve \\
     Variability Index & Ratio of the mean of the square of successive differences & Magntiude &  \cite{Kim_2011}\\
     & to the variance of data points \\
    \hline
    \end{tabular}
    \end{adjustbox}
    \caption{Features used in this work and the properties of the light curves they represent  }
    \label{tab:featuresappendix1}
\end{table*}

\newpage
\onecolumn
\section{Astronomaly Web Interface}

\begin{figure}
    \centering
    \includegraphics[width=18.0cm]{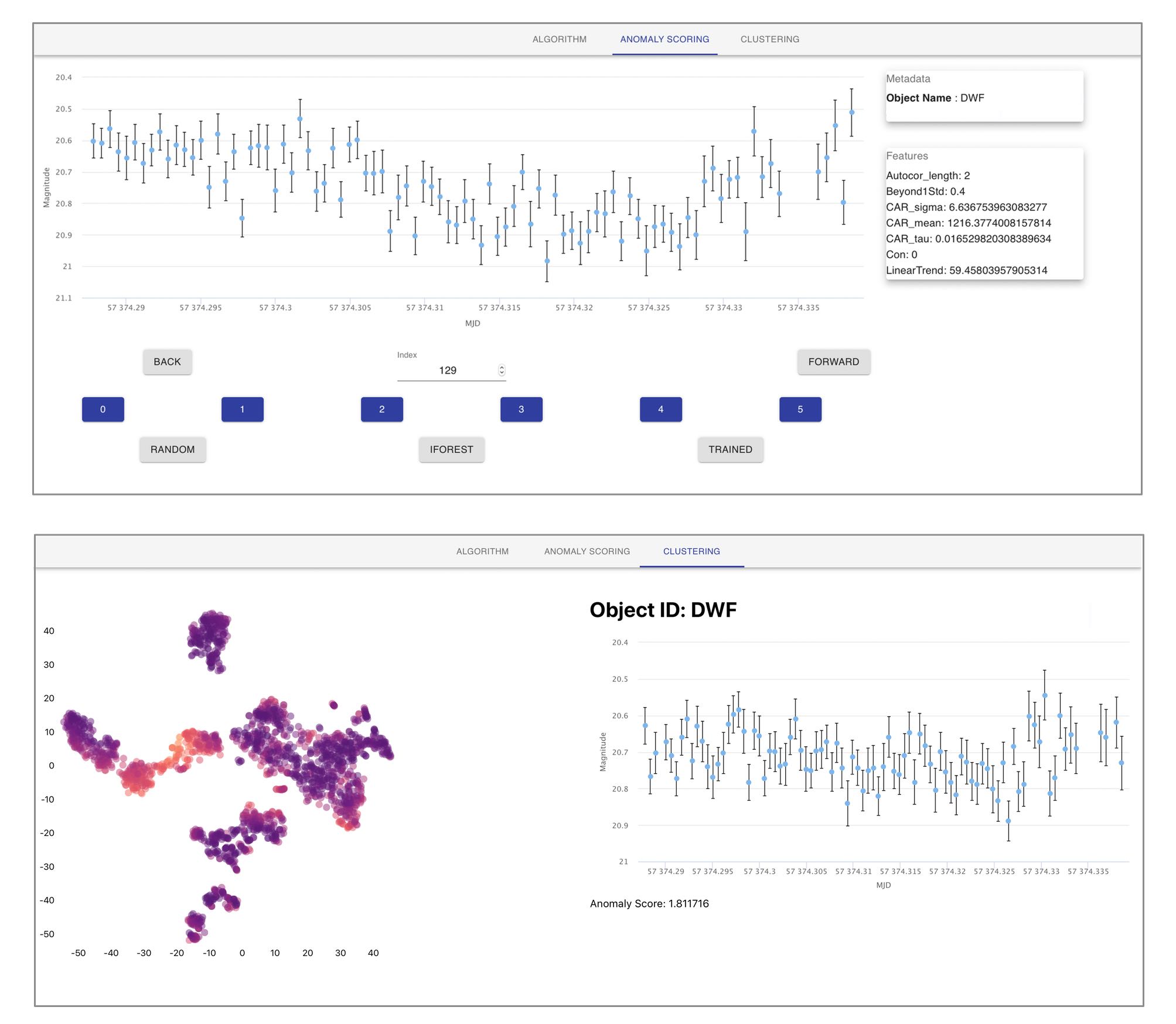}
    \caption{Top) \textit{Astronomaly} web interface `Anomaly Scoring' tab, where light curves can be visually assessed in order of anomaly ranking as determined by the isolation forest algorithm. Bottom) \textit{Astronomaly} web interface `Clustering' tab, displaying an interactive t-SNE plot produced from the input data. The points within the t-SNE can be clicked and then the corresponding light curve will be displayed to the right of the screen. This feature is extremely useful for searching similar light curves based on their features. }
    \label{fig:Astronom}
\end{figure}

\newpage
\onecolumn
\section{Colour Magnitude plot - Newly discovered transients/variables from this work}

\begin{figure}
    \centering
    \includegraphics[height=19.2cm]{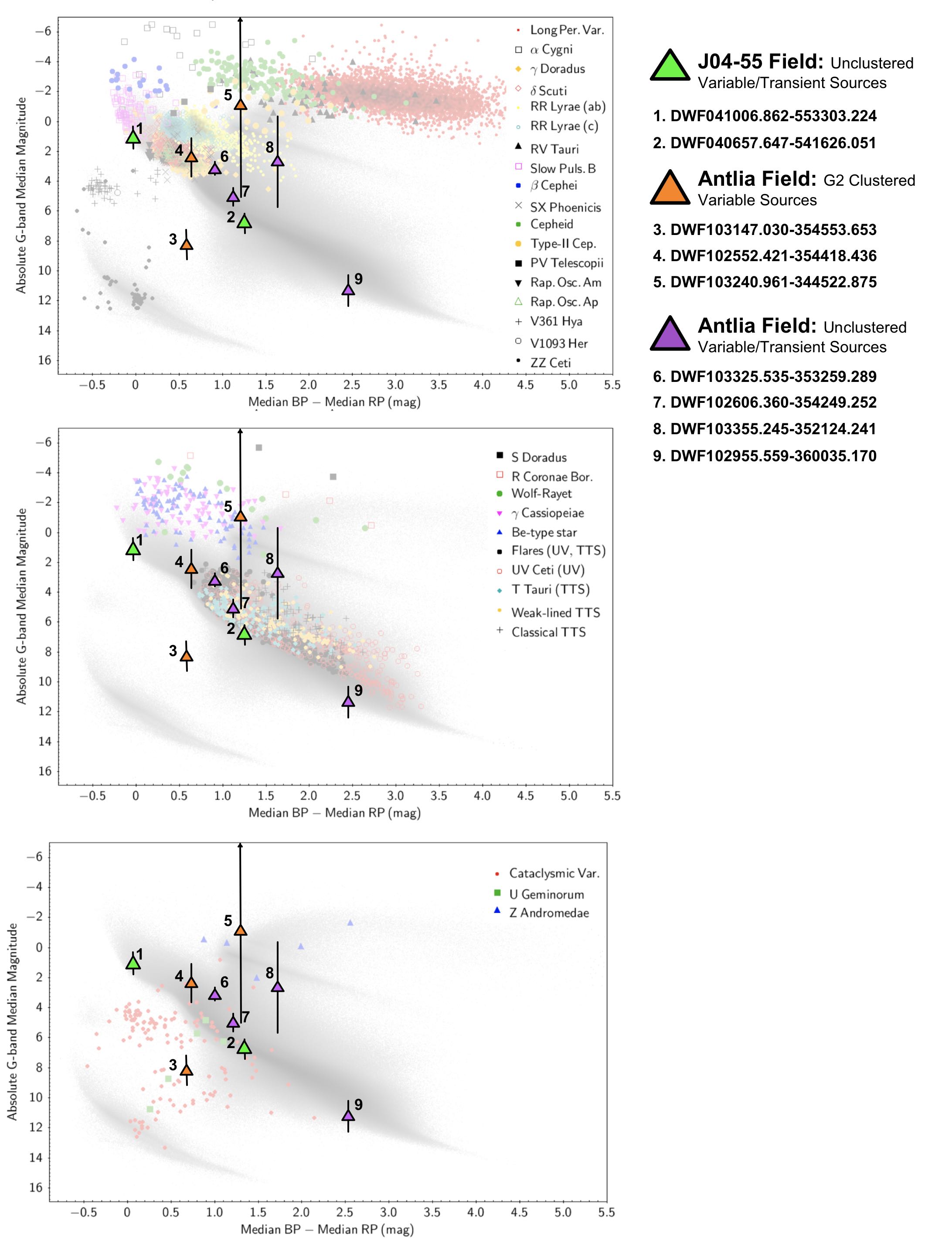}
    \caption{Known pulsating (top panel), eruptive (centre panel), and cataclysmic (bottom panel) variables are shown on the Gaia CMDs \citep{gaiadr2var2019}, with the newly discovered variable and flaring sources (large symbols) overlaid. The green triangles represented sources found in the J04-55 field, the orange represent newly discovered sources from G2 in the Antlia field, and the purple represent the newly discovered sources, which HDBSCAN was not able to cluster. Gaia BP-RP was corrected for galactic reddening  \citep{Schlafly2011}.}
    \label{fig:CMD}
\end{figure} 

\newpage
\onecolumn
\section{Previously catalogued sources}

\begin{table}
    \centering
    \begin{tabular}{cllcl}
     \hline 
    Field  & Catalogue Name & Type & Period (Days) & Notes \\
    \hline 
    \hline
    DWF J04-55 & SSS J041109.9-544851 & W Ursae Majoris eclipsing binary & 0.31 & Identified in this work as anomalous \\
    
    DWF J04-55  & ASAS J040958-5520.2 & Cepheid & 9.20 & Below detection threshold most exposures \\
    DWF J04-55 & ASAS J041436-5441.9 & Contact binary  & 0.45 & Identified in this work as anomalous \\
    
    DWF J04-55 & ASASSN-V J040807.97-541827.2 & W Ursae Majoris eclipsing binary & 0.55  &Identified in this work as anomalous  \\
    
    DWF J04-55 & ASASSN-V J040449.48-552715.9 & W Ursae Majoris eclipsing binary & 0.27 & Identified in this work as anomalous \\
    
    DWF J04-55 & SSS J041229.7-543444 & Asymmetric RR Lyrae & 0.55 &  Below detection threshold most exposures \\
    
    DWF J04-55 & SSS J040348.1-552845 &  W Ursae Majoris eclipsing binary & 0.39 & Below detection thresholdmost exposures \\
    
    DWF J04-55 & SSS J040421.3-551639 & $\beta$ Persei eclipsing binary & 1.15 & Flat light curve, unchanging over observations  \\

    DWF J04-55 & ASAS J040237-5502.5 & Detached eclipsing binary & 1.93 & Flat light curve, unchanging over observations \\
    
    DWF J04-55 & WISE J041127.4-543854 & $\beta$ Persei eclipsing binary & 0.68 & Flat light curve, unchanging over observations  \\
    
    DWF J04-55 & ASASSN-V J041337.83-554819.5 & Variable star of unspecified type & unknown &  Flat light curve, unchanging over observations \\
    
    DWF J04-55 & ASASSN-V J040350.67-545214.6 & Spotted stars that weren't & 0.49 &  Flat light curve, unchanging over observations \\ 
     &&classified into a particular class \\
     
     \hline
    \end{tabular}
    \caption{Variable Star Index (VSX) catalogued variable sources within the DWF J04-55 }
    \label{tab:known}
\end{table}

\begin{table}
    \centering
    \begin{tabular}{cllcl}
     \hline 
    Field  & Catalogue Name & Type & Period (Days) & Notes \\
    \hline 
    \hline
    DWF Antlia & SSS J103047.5-354047 & RR Lyrae & 0.31 & Identified in this work as anomalous \\ 
    DWF Antlia & SSS J102938.8-345416 &  W Ursae Majoris eclipsing binary & 0.27 & Identified in this work as anomalous \\
    DWF Antlia &  SSS J103120.8-354209 &  W Ursae Majoris eclipsing binary & 0.27 & Identified in this work as anomalous \\
    DWF Antlia & ASAS J103038-3558.0 & $\beta$ Persei-type eclipsing binary & 0.72 & Identified in this work as anomalous \\
    DWF Antlia & SSS J103114.5-343834 & RR Lyrae & 0.33 & Identified in this work as anomalous \\
    DWF Antlia & SSS J102742.4-343933 & W Ursae Majoris eclipsing binary & 0.27 & Identified in this work as anomalous \\
    DWF Antlia & SSS J103105.8-360742 & W Ursae Majoris eclipsing binary & 0.44 & Identified in this work as anomalous \\
    DWF Antlia & SSS J102641.7-355130 & W Ursae Majoris eclipsing binary & 0.29 & Identified in this work as anomalous \\
    DWF Antlia & SSS J102919.0-355133 &  Spotted stars that weren't & 0.34 &  Identified in this work as anomalous\\ 
     &&classified into a particular class \\
    DWF Antlia & SSS J102615.2-351023 & RR Lyrae & 0.50 & Below detection threshold most exposures  \\
    DWF Antlia & SSS 110101:103109-350150 & Dwarf novae & unknown &  Flat light curve, unchanging over observation \\
    DWF Antlia & SSS J102933.7-354152 & W Ursae Majoris eclipsing binary & 0.29 & Flat light curve, unchanging over observation \\
    DWF Antlia & SSS J103200.4-353401 & W Ursae Majoris eclipsing binary & 0.44 & Flat light curve, unchanging over observation \\ 
    DWF Antlia & SSS J102734.7-353154 & W Ursae Majoris eclipsing binary & 0.40 & Flat light curve, unchanging over observation  \\
    DWF Antlia & SSS J102717.6-353645 & $\beta$ Persei-type eclipsing binary & 0.89 & Flat light curve, unchanging over observation \\
    DWF Antlia & SSS J103425.0-350405 & W Ursae Majoris eclipsing binary & 0.41 &  Flat light curve, unchanging over observation \\
    DWF Antlia & SSS J102712.4-353219 & RR Lyrae & 0.63 & At saturation limit with photometry affected, \\
    DWF Antlia & SSS J103237.3-345913 &   Spotted stars that weren't & 0.30 & At saturation limit with photometry affected,  \\ 
    &&classified into a particular class & \\
    DWF Antlia & SSS J103436.8-352812 & W Ursae Majoris eclipsing binary  & 0.35 &  At saturation limit with photometry affected \\  
    DWF Antlia & SSS J103157.1-351718 & W Ursae Majoris eclipsing binary & 0.32 & Light curve photometricly affected. \\
   DWF Antlia & SSS J103440.2-351511 &  W Ursae Majoris eclipsing binary & 0.31 &  Affected photometry from CCD edge
    \\ &&&&  identified as such in G6. \\
    DWF Antlia & SSS J102906.8-360355 & W Ursae Majoris eclipsing binary & 0.32 & Affected photometry from CCD edge, \\ &&&&  identified as such in G6. \\
    
    \hline
    
    \end{tabular}
    \caption{Variable Star Index (VSX) catalogued variable sources within the DWF Antlia field}
    \label{tab:knownant}
\end{table}

\newpage
\onecolumn
\section{Light curve traits}

\begin{table*}
    \centering
    \begin{tabular}{ccl}
    \hline
    \textbf{Cluster}  & \textbf{Number of}  & \textbf{Notes}   \\
    & \textbf{Light } & \\
    \hline
    \hline
    unclustered  & 2169 &  Light curves with majority non-detections as well as possible variable sources and photometry affected by telescope dithering. \\
    0 &  6   & Only Detected on five point dithers, either beginning, middle or end of observations. \\
    1  &  6   &  Variable Sources.   \\
    2  &  20   &  Sources near edge of ccd resulting in dimming and brightening as the source moves ccd position during observations.   \\
    3    &  30    & Only Detected on five point dithers, either beginning, middle or end of observations.       \\
    4    & 23     & Only Detected on five point dithers, either beginning, middle or end of observations.      \\
    5   & 7     &  First five point dither detections correction issues of 0.1-0.2 mags.     \\     
    6   & 19     &  First five point dither detections correction issues of 0.1-0.2 mags.     \\
    7   &  58    & First five point dither detections correction issues of 0.1-0.2 mags.      \\
    8   &  16    & Bright Sources on ccd extension 30, Issues with correction over the night.     \\
    9   &  10    &   First five point dither detections correction issues of 0.1-0.2 mags.      \\
    10   &  17    &   First five point dither detections correction issues of 0.1-0.2 mags.      \\
    11    &  107    &   First five point dither detections correction issues of 0.1-0.2 mags.      \\
    12   &  23    &  One or more detections affected by Cosmic Rays, pixel faults, etc.    \\
    13   &  5    &  First five point dither detections correction issues of 0.1-0.2 mags.     \\
    14    &  14    &  Sources near edge of ccd resulting in dimming and brightening as the source moves ccd position during observations.
    \\
    15    & 11     &  Bright Sources on ccd extension 30, Issues with correction over the night.    \\
    16   &  22    &   Bright Sources on ccd extension 30, Issues with correction over the night.   \\   
    17    & 23     &  Sources near edge of ccd resulting in dimming and brightening as the source moves ccd position during observations.    \\
    18    &  20    &  First five point dither detections correction issues of 0.1-0.2 mags.     \\
    19    &  8    &  Sources on ccd extension 30, Issues with correction over the night.    \\
    20    &  226    &  One or more detections affected by Cosmic Rays, pixel faults, etc and faint sources at detection threshold.     \\
    21    &  12    &  Only Detected on five point dithers, either beginning, middle or end of observations.     \\
    22    &  6    &  Only Detected on five point dithers, either beginning, middle or end of observations.     \\
    23    &  7    &  Only Detected on five point dithers, either beginning, middle or end of observations.     \\
    24    &  12    & Sources near edge of ccd resulting in dimming and brightening as the source moves ccd position during observations.
     \\
    25    &  17    &  Defuse or blended sources.     \\
    26    &  156    &  Sources near edge of ccd resulting in dimming and brightening as the source moves ccd position during observations.
    \\
    27    &  11    &  Only Detected on five point dithers, either begining, middle or end of observations.     \\
    28    &  11    &  Only Detected on five point dithers, either begining, middle or end of observations.     \\
    29    &  951    &  Sources near edge of ccd resulting in dimming and brightening as the source moves ccd position during observations.
    \\
    30    &  19    &  Faint sources behind defuse galaxies/ blended point sources     \\
    31     &  5    &  One or more detections affected by Cosmic Rays, pixel faults, etc.    \\
    32    &  5    &   Only Detected on five point dithers, either begining, middle or end of observations.    \\
    33     &  12    & Faint sources at detection threshold.      \\
    34     &  6    & Faint sources at detection threshold.        \\
    35    &   5   &  Faint sources at detection threshold.       \\
    36    &   58279   & Steady light curves.      \\
      \hline
    \end{tabular}
    \caption{Clusters Identified from Antlia field light curves using HDBSCAN.}
    \label{tab:antliaclustersexpanded}
\end{table*}


\bsp	
\label{lastpage}
\end{document}